\documentclass{easychair}

\usepackage{cite}
\usepackage{amsmath,amssymb,amsfonts}
\usepackage{algorithmic}
\usepackage{graphicx}
\usepackage{textcomp}
\usepackage{xcolor}
\usepackage{authblk}
\usepackage{float}
\usepackage{multirow}
\usepackage{makecell}
\usepackage{balance}
\usepackage{url}
\usepackage{longtable}
\usepackage{comment}
\usepackage{soul}

\newcounter{auxFootnote}
\newcounter{auxFootnote1}

\begin{document}

\title{Credulous Users and Fake News: a Real Case Study on the Propagation in Twitter
}


\author{
Alessandro Balestrucci\inst{1,2} \and 
Rocco De Nicola\inst{1,3}
}

\institute{
IMT School for Advanced Study Lucca, Lucca, Italy\\\email{rocco.denicola@imtlucca.it}
\and
 Gran Sasso Science Institute,
 L'Aquila, Italy\\
 \email{alessandro.balestrucci@gssi.it}
 \and
 {CINI Cybersecurity Lab, Roma, Italy}
 }



\maketitle

\begin{abstract}
Recent studies have confirmed a growing trend, especially among youngsters, of using Online Social Media as favourite information platform at the expense of traditional mass media. Indeed, they can easily reach a wide audience at a high speed; but exactly because of this they are the preferred medium for influencing public opinion  via so-called fake news. Moreover, there is a general agreement that the main vehicle of fakes news are malicious software robots (\textit{bots}) that automatically interact with human users.

In previous work we have considered the problem of tagging human users in Online Social Networks as \textit{credulous} users. Specifically, we have considered credulous those users with relatively high number of bot friends when compared to total number of their social friends. We consider this group of users worth of attention because they might have a higher exposure to malicious activities and they may contribute to the spreading of fake information by sharing dubious content.

In this work, starting from a dataset of fake news, we investigate the behaviour and the degree of involvement of \textit{credulous} users in fake news diffusion. The study aims to: (i)  fight fake news by considering the content diffused by \textit{credulous} users; (ii) highlight the relationship between \textit{credulous} users and fake news spreading; (iii) target fake news detection by focusing on the analysis of specific accounts more exposed to malicious activities of bots. Our first results demonstrate a strong involvement of \textit{credulous} users in fake news diffusion. This findings are calling for tools that,  by performing  data streaming  on credulous' users actions, enables us to perform targeted fact-checking. 
\\\\
\textbf{keywords} -- 
Social Networks Analysis and Mining, Fake News, Twitter, Credulous  Users

\end{abstract}

\section{Introduction}
\label{sec:1Intro}
Pervasiveness and ease of use of Online Social Media (OSM), like Twitter and Facebook, have lead to new ways for people to keep up with news and for their propagation. 
Recent studies~\cite{aneez2019reuters,nic2018reuters,newman2017reuters} confirm the growing trend  of using digital media as the favourite source of information, especially among youngsters~\cite{newman2016digital}. By routinely checking pages of interest and channels, people get informed effortlessly. With OSM, publishers reach larger audience than with traditional mass media, like newspapers, radio, etc., and news are disseminated easily and at a faster rate. This has however brought concerns about the veracity of the news circulating on OSM. Fake news and mis/disinformation~\cite{lazer2018science} are well known problems, against which both government~\cite{andorfer2017spreading}, researchers~\cite{shu2017fake} and social media administrators\footnote{Facebook: \url{https://tinyurl.com/y3yzvpah}}$^,$\footnote{Twitter: \url{https://tinyurl.com/y3efs8s5}} are struggling. 

Several approaches have been developed to counteract the spread of this phenomenon. Some of them aim at detecting  \textit{bots}~\cite{ferrara2016rise}, i.e., automated accounts interacting with human users; others use natural language processing (NLP) techniques~\cite{oshikawa2018survey} by analyzing  the actual contents of messages. In this work, we look at fake news from a different perspective, aiming to figure out the extent of relationships between fake news and \textit{credulous} users~\cite{balestrucci2019identification}. The latter are human-operated accounts with a relatively  high percentage  of  \textit{bot-followees} among the total number of their followees and are thus more exposed to bots' malicious activities. 
The motivation for studying the phenomenon from this perspective is twofold, indeed credulous users: (i)  may unconsciously contribute to fake news dissemination~\cite{DBLP:conf/ideal/BalestrucciNPT19,efthimion2018supervised};
(ii) may be affected by  malicious activities that are more effective when performed on a targeted audience~\cite{bovet2019influence,albright2017welcome}.

Starting from a publicly available dataset of fake news which concerned with politics and gossips~\cite{shu2018fakenewsnet} published on Twitter, we have studied the involvements of \textit{credulous} users  in terms of the number of their tweets with fake news they posted and the number of credulous that have shared fakes. By jointly exploiting bot detection and credulous detection approaches~\cite{DBLP:conf/ideal/BalestrucciNPT19}, we have seen that: (i) \textit{credulous} users produce more fake news tweets than not-credulous ones; (ii) \textit{credulous} users  publish less real news than fake ones and (iii) the extent to which \textit{credulous} publish fake news depends on the topic.


This findings are calling for tools that, by performing  data streaming  on credulous' users actions, enables us to perform targeted fact-checking. A possible exploitation of this could be a system that, by attentioning credulous users performs on line data streaming by "listening" their activities (e.g., tweets, retweets, replies, mentions, etc.) in real time. As soon as a credulous user publishes on his social dashboard, the tool could analyze the content, with, e.g., text mining and/or NLP techniques, the reliability of the source. 
Obviously, to reduce the amount of content to inspect, a key component of the system is the part concerned with credulous detection that reduces the number of human-operated accounts to investigate. 
This can be made more efficient by setting up a priority inspection measure, based on the content production rate of credulous users, to firstly focuses on the more active ones.

The rest of the paper is structured as follows: Section \ref{sec:2RW} considers related works; 
Section \ref{sec:3ExpSet} describes the used dataset and the approach we have followed;
Section \ref{sec:4ExpRes} illustrates the experimental results; Section \ref{sec:5Disc} discusses the main findings, and Section \ref{sec:6Conc} concludes the paper.
\section{Related Work}
\label{sec:2RW}
This paper aims at investigating the relationship between \textit{credulous} users and fake news; in particular, the way the former behave in OSM and the extent to which they are involved in spreading the latter using tweets. 
Rather than providing a complete literature review about the many approaches to fake news detection, we concentrate just on work investigating on the way bots' activities influence humans in OSM. For comprehensive surveys on  fighting fakes, we refer the reader to other  papers~\cite{shu2017fake,bondielli2019survey}.

An interesting study misinformation spreading in OSM is presented in~\cite{shao2017spread}. The authors analyzed 14 million of tweets published during 2016 U.S. presidential campaign and found non-negligible evidence of the role that bots have had in fake news dissemination. Furthermore, they identified specific vulnerability of humans 
that had been retweetting malicious bots; thus actively contributing to fake news spreading.






The notion of \textit{susceptible} users, i.e., OSM's users who interact with bots, is introduced in~\cite{wagner2012social}.  
A binary classifier was trained to single out \textit{susceptible} users; and the need of introducing  protection mechanisms in Social Media to prevent support of malicious activities by human-operated accounts was advocated. 
%
Human-operated accounts ``attacked'' by bots through tweet mention are studied in~\cite{wald2013predicting} with the aim of predicting whether an interaction will start. 
Instead, we consider those human users who have already started interacting with bots to discover the extent of their involvement in spreading of fake news.

A live lab experiment testing the ability of social bots 
to shape or influence the social graph of human-operated accounts in Twitter is presented  in~\cite{mitter2014categorization} and it is  observed that the activities of social bots can affect the links established between targeted users. 

Starting from the definition of \textit{susceptible} users~\cite{wagner2012social} and with aims close to ours, the notion of \textit{gullible} users and their susceptibility to fake news is introduced in~\cite{DBLP:conf/websci/ShenCGLYL19}. Five degrees of susceptibility (referred to a user's reply to a fake news) are defined and a multi-class classifier is trained to predict users susceptibility level. The classifier achieves an AUC of 0.82. Indeed, their aim is similar to ours, but we do not rely on users replies to deem them as \textit{credulous}, but on the number of bots they follow, 
and our objective is not a classification, but to find evidence of their involvement 
in fake news spreading.


Beside the definitions of \textit{susceptible} users~\cite{wagner2012social} and \textit{gullible} users~\cite{DBLP:conf/websci/ShenCGLYL19}, we mention our previous works~\cite{balestrucci2019identification,DBLP:conf/ideal/BalestrucciNPT19,balestrucci2020botfollow} where \textit{credulous} users have been studied.
In~\cite{balestrucci2019identification} the concept of \textit{credulous} users has been introduced and heuristics 
have been used to label as credulous human-operated accounts; humans are ranked by relying on the ratio between the number of bots (recognized  by a bot detector) they follows and their total number of followees\footnote{1st dataset of credulous users: \url{https://tinyurl.com/y6lod2yz}}. 
A method for automatically identifying \textit{credulous} users, has been introduced in~\cite{DBLP:conf/ideal/BalestrucciNPT19} for the first time. 
In this work, classification performance achieved an accuracy of 93.27\% and an \textit{AUC} (Area Under the ROC curve) of 0.93.
From a perspective of regression, rather than for classification purposes,  the goal of the work in~\cite{balestrucci2020botfollow} has been to single out a technique for quantifying the percentage of \textit{bot-followees}, i.e., bots followed by human-operated accounts in Twitter. The best regression model is trained with the \textit{SMOreg}~\cite{shevade2000improvements} algorithm, achieving a Mean Absolute Error of 3.62\%. 
The current paper differs in scope and strategy with our previous work. Its primary objective is to investigate the way \textit{credulous} users deal with fake news and to understand whether analyzing \textit{credulous} users profiles can be useful to support other fake news detection techniques. In this case, the basic strategy relies on observing tweets containing fake news and their spreaders, without considering any longer regression or classification.

\section{Experimental Setup}
\label{sec:3ExpSet}

\subsection{Dataset}
\label{subsec:data}
We take as starting point a publicly available dataset of fake news, called \textit{FakeNewsNet}\footnote{FakeNewsNet Dataset: \url{https://tinyurl.com/uwadu5m}}~\cite{shu2017fake,shu2017exploiting,shu2018fakenewsnet}. For each item the following information is provided: a unique identifier (\textit{id}), the publisher (in \textit{url} form), the content of the news (\textit{title}), a list tweets (as Twitter \textit{ids}) containing the news and the information about its ``veracity'' (fake or real). 
To label the news, the authors in~\cite{shu2018fakenewsnet} used two fact-checking websites: \textit{PolitiFact}\footnote{\url{https://www.politifact.com/}} and \textit{GossipCop}\footnote{\url{https://www.gossipcop.com/}}. In the former, fact-checking 
was performed by politics experts (e.g., journalists) whom labelled news as fake or real. 
In the latter, a numerical scores was assigned to news to indicate reliability, ranging from 0 (fake) to 10 (real). 

\newcommand{\STAB}[1]{\begin{tabular}{@{}c@{}}#1\end{tabular}}

\begin{table}[!htbp]
	\begin{centering}
	\caption{FakeNewsNet Dataset: original and retrieved content}	\begin{tabular}{ccccccc}
		\cline{1-7}
		\multirow{2}{*}{}
        & \makecell[c]{\\[-1mm]~}&\multicolumn{2}{c}{\textbf{News}}&&\multicolumn{2}{c}{\textbf{Tweets}}\\
		\cline{3-4}\cline{6-7}
		& \makecell[c]{\\[-1mm]~} & Original & Retrieved && Original & Retrieved \\
		\hline
		\multicolumn{1}{c}{\multirow{2}{*}{\STAB{\rotatebox[origin=c]{90}{Politic~~}}}} &  \multicolumn{1}{c}{\STAB{\rotatebox[origin=c]{90}{~Fake~}}} & 432 & 392 && 165,356 & 141,421\\
		\cline{2-7}
		\multicolumn{1}{c}{}& \multicolumn{1}{c}{\STAB{\rotatebox[origin=c]{90}{~Real~}}}&622 & 407&& 417,072& 357,655\\ \hline
		\multicolumn{1}{c}{\multirow{2}{*}{\STAB{\rotatebox[origin=c]{90}{Gossip}}}}& \multicolumn{1}{c}{\STAB{\rotatebox[origin=c]{90}{~Fake~}}} & 5,323& 5,135&& 598,299& 518,502\\
		\cline{2-7}
		\multicolumn{1}{c}{}& \multicolumn{1}{c}{\STAB{\rotatebox[origin=c]{90}{~Real~}}}& 16,817& 15,759&& 881,627&812,719 \\ \hline\\
		\end{tabular}
		
		\label{tab:Data}
	\end{centering}
	\vspace{-1em}
\end{table}

Table~\ref{tab:Data} outlines the structure of the dataset. 
The original dataset contained 432 fake and 622 real political news (see in row \textit{Politic} the column \textit{Original}). 
However, on Twitter we were only able to find tweets of 392 fake and 407 real news (column headed \textit{Retrieved}). 
The number of \textit{tweets} (column titled \textit{Tweets}) containing such news was initially of 165,356 on fake and 417,072 on real news; but we could find  only 141,421 tweets related to fake news and 357,655 tweets related to real ones.
The numerical mismatch between the \textit{original} data and the \textit{retrieved} data is almost certainly due to  deletion. 
Regarding the other topic, (row headed \textit{Gossip}), out of 5,323 fake and 16,817 real news, we got 5,135 and 15,759 news respectively. And it was possible to find a total of 1,331,221 tweets; 518,502 related to fake news and 812,719  containing real news. Obviously, in our study, we will only use retrieved data in both cases.

\subsection{Experimental Approach}
\label{subsec:expApp}
To pursue the goals of this work, we single out three sequential tasks: 
(i) tweets' authors identification, 
(ii) distinction between automated (bots) and human-operated authors, and 
(iii) distinction between \textit{credulous} and \textit{not-credulous} users among the human-operated authors. 

\subsubsection{Tweets' authors identification}
\label{subsubsec:author}
In this phase, we aim to identify  the Twitter accounts that published the tweets listed in \textit{FakeNewsNet} dataset, and thus their authors. 
Starting from the tweets' id and using Twitter API\footnote{Twitter API libraries: \url{https://tinyurl.com/rfte3k2}}, we collected 1,731,422 
tweets out the \textit{original} list of almost 2 million. 
It might be worth noting that some tweets contain more than one news, and thus
some tweets are counted more than once in Table~\ref{tab:Data}. This explains the numerical mismatch between the collected tweets and retrieved ones (value's sum of 4th column). 

At the end of this phase, in addition to tweets' data, we collected the profile's data
\footnotemark\setcounter{auxFootnote}{\value{footnote}} of 536,513 Twitter accounts which are their authors. 
\footnotetext{Twitter User Object: \url{https://tinyurl.com/y5s5kpuw}}

\subsubsection{Bot detection}
\label{subsubsec:bot}
The goal of this phase is to distinguish, among the set of authors obtained in the previous phase, between human-operated accounts and bots.
To this purpose, we used a bot detector (i.e., a decision model able to recognize automated accounts) introduced in our previous work~\cite{DBLP:conf/ideal/BalestrucciNPT19}. 
The classification model is based on Random Forest~\cite{breiman2001random}, achieving an \textit{accuracy} (instances correctly classified) of 98.41\% and an area under the ROC curve (AUC) of 1.00. It relies on a set of 30 features. Specifically, such features are obtained by combining the feature sets of \textit{Botometer+} and \textit{ClassA-}. The reader is referred to~\cite{balestrucci2020botfollow} for details. 
For each users, \textit{Botometer+} considers the \textit{timeline} (the list of published tweets) and its \textit{mentions} (the tweets that mention the user). \textit{ClassA-}'s, instead, relies on users' profile data to determine their features\footnotemark[\value{auxFootnote}]. 

It is worth to notice that to retrieve the \textit{Botometer+}'s features is needed a connection to a web service\footnote{Botometer~\cite{Varol17}: \url{https://tinyurl.com/yytf282s}} 
and to classify tweets' authors it is mandatory to obtain features from both \textit{Botometer+} and \textit{ClassA-}. 
In this way, 479,569 authors have been classified as human-operated.
%
%

\subsubsection{Credulous classification}
\label{subsubsec:credulous}
The last task aims at singling out \textit{credulous} users among the human-operated authors. To this purpose a refined version of the approach presented in~\cite{DBLP:conf/ideal/BalestrucciNPT19}, has been adopted. In~\cite{DBLP:conf/ideal/BalestrucciNPT19} the ground-truth was strongly imbalanced (316 credulous vs. 2,522 not-credulous). 
There, the set of \textit{not-credulous} users has been divided into smaller subsets, by a randomly instance selection (without replacement), and each subset has cardinality equal to the number of \textit{credulous} users. Each subset of \textit{not-credulous} users has been then merged with the \textit{credulous} instances, creating eight new ``sub-datasets'' called \textit{folds}; differing from each other for \textit{not-credulous} instances only. Then, for each \textit{fold}, several decision models have been trained and cross-validated. The fold \textit{OneR}\footnotemark\setcounter{auxFootnote1}{\value{footnote}} turned out to be the best performing one.
\footnotetext{OneR~\cite{holte1993very}: \url{https://tinyurl.com/wgpezcp}}
%
%
But for some folds, other algorithms (specifcally \textit{JRip}\footnote{JRip~\cite{Quinlan1993}: \url{https://tinyurl.com/ufke2zb}} and \textit{RepTree}\footnote{RepTree~\cite{cohen1995fast}: \url{https://tinyurl.com/qkl8nko}}) turned out to produce better models (see \textit{fold} 4 and 5 in Table~\ref{tab:8credclass}). 

Instead of classifying authors using the best classifier (which has been trained on just one fold in~\cite{DBLP:conf/ideal/BalestrucciNPT19}), for labelling authors we use all eight \textit{credulous} classifiers in Table~\ref{tab:8credclass}.
\\
\begin{table}[htbp]
	\begin{center}
		\begin{tabular}{clccccccc}
		\hline
 & & \multicolumn{6}{c}{\textit{evaluation metrics}}\\
\cline{3-8}
 \textit{Fold} & \textit{alg} & \multicolumn{1}{r}{\textit{accuracy}} & \multicolumn{1}{r}{\textit{prec.}} & \multicolumn{1}{r}{\textit{rec.}} & \multicolumn{1}{r}{\textit{F1}} 
& \multicolumn{1}{r}{\textit{MCC}} & \multicolumn{1}{r}{\textit{AUC}} \\
\hline
1 & OneR & \multicolumn{1}{r}{98.26} & \multicolumn{1}{r}{0.98} & \multicolumn{1}{r}{0.98} & \multicolumn{1}{r}{0.98} & \multicolumn{1}{r}{0.97} & \multicolumn{1}{r}{0.98} \\
2 & OneR & \multicolumn{1}{r}{95.73} & \multicolumn{1}{r}{0.96} & \multicolumn{1}{r}{0.96} & \multicolumn{1}{r}{0.96} & \multicolumn{1}{r}{0.92} & \multicolumn{1}{r}{0.96} \\
3 & OneR & \multicolumn{1}{r}{94.15} & \multicolumn{1}{r}{0.95} & \multicolumn{1}{r}{0.94} & \multicolumn{1}{r}{0.94} & \multicolumn{1}{r}{0.89} &  \multicolumn{1}{r}{0.94} \\
4 & JRip & \multicolumn{1}{r}{90.67} & \multicolumn{1}{r}{0.92} & \multicolumn{1}{r}{0.91} & \multicolumn{1}{r}{0.91} & \multicolumn{1}{r}{0.83} &  \multicolumn{1}{r}{0.89} \\
5 & RepTree & \multicolumn{1}{r}{91.93} & \multicolumn{1}{r}{0.93} & \multicolumn{1}{r}{0.92} & \multicolumn{1}{r}{0.92} & \multicolumn{1}{r}{0.85} &  \multicolumn{1}{r}{0.90} \\
6 & OneR & \multicolumn{1}{r}{90.35} & \multicolumn{1}{r}{0.92} & \multicolumn{1}{r}{0.90} & \multicolumn{1}{r}{0.90} & \multicolumn{1}{r}{0.82} &  \multicolumn{1}{r}{0.90} \\
7 & OneR & \multicolumn{1}{r}{90.93} & \multicolumn{1}{r}{0.92} & \multicolumn{1}{r}{0.91} & \multicolumn{1}{r}{0.91} & \multicolumn{1}{r}{0.83} &  \multicolumn{1}{r}{0.91} \\
8 & OneR & \multicolumn{1}{r}{96.65} & \multicolumn{1}{r}{0.97} & \multicolumn{1}{r}{0.97} & \multicolumn{1}{r}{0.97} & \multicolumn{1}{r}{0.93} &  \multicolumn{1}{r}{0.97} \\
\hline
  &  &  &  &  & \\
		\end{tabular}
\caption{The eight Credulous Classifiers}
		\label{tab:8credclass}
	\end{center}
	
\end{table}
Specifically, the basic idea is that each author has to be classified by means of the classifier trained on the \textit{fold} most ``similar'' to it. Hence, for each human author singled out in Section~\ref{subsubsec:bot}, the distance between author's feature representation and the centroids of each \textit{fold} is computed. The author is then classified using the classifier trained on the fold whose centroid is closest to it. Classifiers selection is performed with a specific tool we have implemented; and 350,622 human authors have been classified as \textit{credulous} users.

\subsection{Investigation targets}
\label{subsec:invTar}
At this stage, we have all the information to study the relationships between fake news and \textit{credulous} users. In particular, we want to investigate potential relationships under three different perspectives, aiming at understanding whether \textit{credulous} users do significantly contribute to fake news production and/or propagation in Twitter. First, 
we want to look for  numerical differences between the amount of fake and real news produced/diffused by the three categories of user/author (namely, \textit{credulous}/\textit{not-credulous} and bots) and on both news' topic. 
Second, we want to compare the quantity of fake/real tweets, i.e., the tweets containing a fake/real news,
by contrasting first bots and humans and then \textit{credulous} and not-credulous. 
To avoid that the numerical unbalancing between fake and real tweets (e.g., see in Table~\ref{tab:Data} political \textit{retrieved} tweets) may lead to inaccurate observations 
we will also consider a randomly selected subset, among the set of real-news tweets, with the same number of fake-news tweets. 
Third, we want to quantify the \textit{authors' level of involvement} in fake news spreading/production by counting, for each category, how many of them are authors of tweets containing: at least one fake news, at least one real news, only fake news and only real news. 

\section{Experimental Results}
\label{sec:4ExpRes}

In this section we present the experimental results, 
%
by relying on the approach described in Section~\ref{subsec:expApp}. We start by showing the results obtained for bot and \textit{credulous} detection in Table~\ref{tab:botDetector} (Sections~\ref{subsubsec:bot} and~\ref{subsubsec:credulous}). 
\begin{table}[!htbp]
\begin{center}

\begin{tabular}{llcccc}
		\cline{2-5}
		&\makecell[c]{\\[-1mm]~}& \textbf{Politc} & \textbf{Gossip} & \textbf{Union} & \\[0.2mm]
		\cline{2-5}
		\makecell[c]{\\[-0.5em]~}&\#Bot & 27,137 & 34,160 & 56,548 & \\
		&\#Human& 256,561 & 247,113 & 479,569 & \\
		\cline{2-5}
		\makecell[c]{\\[-0.5em]~}&\#Credulous& 185,196 & 178,715 & 350,622 & \\
		&\#Not-Credulous & 71,365 & 68,398 & 128,947 & \\
	\cline{2-5}\\
\end{tabular}
\label{tab:botDetector}
\caption{Bot Detector and Credulous detector outcomes}
\end{center}
\end{table}

In the first column there are the different types of users; in the first macro-row the difference is based on the ``automation'' of an account, the second macro-row reported the numbers of human-operated accounts labeled as \textit{credulous} or \textit{not-credulous} users. Each column reports the number of users which tweeted about a certain topic. 
We could not classify 396 accounts, because the \textit{Botometer} web service did not return features.

Table~\ref{tab:TopicXUsers} reports the amount of news for each category of users, i.e., credulous/not-credulous users and bots (1st column), 
for each topic (\textit{Politic} or \textit{Gossip}). The 2nd and the 3rd column, (macro-column headed \textit{Politic}), indicate the amount of political news, respectively fake and real, that users have  used in their tweets. The 4th and 5th column, instead, indicate the number of real and fake news about the gossip, respectively.

\begin{table}[htbp]

	\begin{center}
		\begin{tabular}{lccccc}
		\cline{1-6}
		\multirow{2}{*}{\textbf{}} \makecell[c]{\\[-2mm]~}& \multicolumn{2}{c}{\textbf{Politic}} && \multicolumn{2}{c}{\textbf{Gossip}}\\
		\cline{2-3}\cline{5-6}
		\makecell[c]{\\[-2mm]~}& \textit{Fake} & \textit{Real}&& \textit{Fake} & \textit{Real} \\
		\hline\\[-2mm]
		Credulous & 373 & 364 && 4,121 & 14,486 \\
		NotCredulous & 361 & 366 && 4,768 & 13,418\\
		Bot & 350 & 332 && 4,470 & 15,050 \\
		\hline\\
		\end{tabular}
		\caption{Users' topic coverage by their tweets}
		\label{tab:TopicXUsers}
	\end{center}
\end{table}

With their tweets, credulous users cover 373 and 364 fake and real political news respectively.
The number of news covered by \textit{not-credulous} users is 361 fake and 366 real political news. For the sake of completeness, the same information is reported for bots too; the amount of news covered by them is 350 fake and 332 real political facts. 

When considering \textit{Gossip} news, starting from a retrieved set of 5,135 fake news, we have the following numbers: 4,121 by \textit{credulous} users, 4,768 by \textit{not-credulous} users and 4,470 by automated accounts. For real news, we retrieved 15,759 tweet, of which 14,486 come from \textit{credulous} users, 13,418 from \textit{not-credulous} and 15,050 from bots.

Table~\ref{tab:TwPolitic} and~\ref{tab:TwGossip} provides a more detailed view of our experiments.
For each category of users, the number of tweets containing fake news (column \textit{FN}) and real news (column \textit{RN}) for both politics (Table~\ref{tab:TwPolitic}) and gossip (Table~\ref{tab:TwGossip}) news is reported. In both tables, the 4th column (called \textit{RN$_{rnd}$} ) and 5th column (called \textit{RN*$_{rnd}$}) are introduced to mitigate the unbalance between fake and real tweets, in accordance with the discussion in Section~\ref{subsec:invTar}. With \textit{RN$_{rnd}$} we denote a subset of \textit{RN} whose entries have been randomly selected, from the \textit{original} list of tweets of Table~\ref{tab:Data}, in order to have $|$\textit{RN$_{rnd}$}$| = |$\textit{FN}$|$. While in \textit{RN*$_{rnd}$}, tweets are taken from the \textit{retrieved} list of Table~\ref{tab:Data}
($|$\textit{RN*$_{rnd}$}$| = |$\textit{FN$(Tot.)$}$|- |$\textit{FN$(n.a.)$}$|$).

\begin{table}[!htbp]
	\begin{center}
		\begin{tabular}{lcccc}
		\hline
		\\[-2mm]& FN & RN & RN$_{rnd}$ & RN$^*_{rnd}$\\[1mm]\hline
		\\[-2mm]Tot. & 165,356 & 417,072 & 165,356 & 141,421\\
		Bot & 19,888 & 45,924 & 18,120 & 18,013 \\
		n.a. & 23,935 & 59,417 & 23,519 & 0 \\
		Human & 121,533 & 311,731 & 123,717 & 123,408 \\
		\hline 
		\\[-2mm]Credulous & 84,362 & 197,454 & 78,528 & 77,994 \\
		Not Credulous & 37,171 & 114,277 & 45,193 & 45,414 \\
		\hline\\
		\end{tabular}
		\caption{Number of tweets about political fact}
		\label{tab:TwPolitic}
 	\end{center}
\end{table}

 \begin{table}[H]
 	\begin{center}
		\begin{tabular}{lcccc}
		\hline
		\\[-2mm]& FN & RN & RN$_{rnd}$ & RN$^*_{rnd}$\\[1mm]\hline
		\\[-2mm]Tot. & 598,299 & 881,627 & 598,299 & 518,502\\
		Bot & 116,398 & 486,907 & 330,425 & 310,810 \\
		n.a. & 79,797 & 68,908 & 46,552 & 0 \\
		Human & 402,104 & 325,812 & 221,322 & 207,692 \\
		\hline 
		\\[-2mm]Credulous & 244,690 & 209,579 & 142,246 & 133,518 \\
		Not Credulous & 157,414 & 116,233 & 79,076 & 74,174 \\
		\hline\\
		\end{tabular}
		\caption{Number of tweets about gossip fact}
		\label{tab:TwGossip}
	\end{center}
\end{table}

Table~\ref{tab:TwPolitic} presents the information related to the tweets on political fact. Almost 20k fake tweets have been produced by bots, more than 121k fake tweets have been produced by human-operated accounts; while for 24k fake tweets it has not been possible to retrieve their information from Twitter 
(row named \textit{n.a.}). When considering \textit{human} users, we can see that the number of fake tweets (\textit{FN}) published by \textit{credulous} users (i.e., 84k) overcomes the number of those published by \textit{not-credulous} ones (i.e., 37k).
\begin{table}[!htbp]
	\begin{center}
		\hspace*{-2cm}\begin{tabular}{lccccccccccc}
		\cline{1-12}
		\multirow{2}{*}{}
        \makecell[c]{\\[-1mm]~}&\multicolumn{3}{c}{\textbf{Credulous}}&&\multicolumn{3}{c}{\textbf{NotCredulous}}&&\multicolumn{3}{c}{\textbf{Bot}}\\\cline{2-4}\cline{6-8}\cline{10-12}
		\makecell[c]{\\[-1mm]~}& \multicolumn{1}{c}{\#Users} & Average & St.Dev. && \#Users & Average & St.Dev. && \#Users & Average & St.Dev. \\\hline
		\#Fake News$\geq1$ & \multicolumn{1}{r}{54,828} & 1.54 & 2.79 &&19,525 &1.90 &~5.05 &&9,622 & 2.07&4.53\\
		\#Real News$\geq1$ & \multicolumn{1}{r}{138,113} & 1.43 &2.91 && 57,839& 1.98& 10.22&& 45,924& 2.36& 9.11\\
		Only Fake News & \multicolumn{1}{r}{47,083} & 1.40&2.15 &&13,526 &1.60 &~4.61 && 7,658&1.79 &3.78 \\
		Only Real News & \multicolumn{1}{r}{130,368} & 1.37&2.53 && 51,480&1.78 &10.25 &&17,515 &2.15 &8.80 \\
		\hline\\
		\end{tabular}
		\caption{Users that tweetted in political topic}
		\label{tab:ActUsrsPol}
	\end{center}
\end{table}

\begin{table}[!htbp]
	\begin{center}
		\hspace*{-2cm}\begin{tabular}{lccccccccccc}
		\cline{1-12}
		\multirow{2}{*}{}
        \makecell[c]{\\[-1mm]~}&\multicolumn{3}{c}{\textbf{Credulous}}&&\multicolumn{3}{c}{\textbf{NotCredulous}}&&\multicolumn{3}{c}{\textbf{Bot}}\\\cline{2-4}\cline{6-8}\cline{10-12}
		\makecell[c]{\\[-1mm]~}& \multicolumn{1}{c}{\#Users} & Average & St.Dev. && \#Users & Average & St.Dev. && \#Users & Average & St.Dev. \\\hline
		\#Fake News$\geq1$ & \multicolumn{1}{l}{147,158} & 1.66& 8.01&&56,451 &2.79 &33.56 &&25.818 &4.51 &22.00\\
		\#Real News$\geq1$ &\multicolumn{1}{l}{\hspace{0.5em}39,528} & 5.30& 143.59&& 19,047& 6.10& 88.26&&14,620 &33.30 &606.39 \\
		Only Fake News &\multicolumn{1}{l}{139,187} &1.45 &6.97 &&49,351 &1.81 &6.51 &&19,540 &2.39 &12.79\\
		Only Real News &\multicolumn{1}{l}{\hspace{0.5em}31,557} &1.35 & 2.96&& 11,947& 1.52& 4.72&&\hspace{0.5em}8,342 &2.19 &10.95 \\
		\hline
		\end{tabular}
		\caption{Users that tweeted in gossip topic}
		\label{tab:ActUsrsGoss}
	\end{center}
	\vspace{-1em}
\end{table}

Although the tweets in \textit{RN} are more numerous than \textit{FN}, the number of tweets authored by \textit{credulous} overcomes that of not-credulous, but with a lower proportionality (197,454 by \textit{credulous} users and 114,277 by \textit{not-credulous} users). 
But, because the \#\textit{RN} humans' tweets are almost three times the \#\textit{FN} ones, looking to the values related to \textit{RN$_{rnd}$} and \textit{RN*$_{rnd}$} columns can led to a better comparison. 
In fact, we can note that the number of real tweets published by \textit{credulous} users (in 4th and 5th column) are similar and in any case lower than in \textit{FN} (see Table~\ref{tab:TwPolitic}).
%
We preferred to use such subsets, rather than resorting to ratios and percentages, in order to provide direct numerical differences. However, had we used percentages, nothing would have changed. Obviously, we had to take into account that the percentage of \textit{FN} tweets from \textit{credulous} users, out of the total set of human's \textit{FN}-tweets, is 60.85\%, while the percentage tweeted by \textit{not-credulous} users is 39.15\%.

As far as bots are concerned, despite \textit{RN}'s value seems higher then \textit{FN}, the amount of fake and real tweets is more or less the same (see the 4th and 5th column).

Switching to the case of tweets containing gossip news (Table~\ref{tab:TwGossip}), we can immediately notice that, 
like Table~\ref{tab:TwPolitic}, also here there is a superiority in tweet's production by \textit{credulous} users. 
In particular, by focusing on the fake tweets column (\textit{FN}), we can see that even for this topic the amount of tweets published by \textit{credulous} users (244,690) is greater than \textit{not-credulous} (157,414). 
This superiority is confirmed even in all the \textit{RN}'s cases (see 3rd, 4th and 5th columns) but with lower numbers. 
Surprisingly, by looking about bots, they authored a lot of real tweets, precisely 468,907 (\textit{RN}), that represents more than the 50\% of all real tweets; conversely, they published only 116,398 fake tweets (FN), less than the 20\% of \textit{Tot. FN}.

Tables~\ref{tab:ActUsrsPol} and~\ref{tab:ActUsrsGoss}  present the results deriving by looking to what extent, for each topic, the three categories of users (macro-columns' headers in both tables), are participating. 
Specifically, it is reported the amount of users that are authors of: 
at least a fake tweet (1st row, \textit{\#Fake News$\geq1$} ), 
at least a real tweet (2nd row, \textit{\#Real News$\geq1$}), \textit only fake tweets (3rd row, \textit{\#Only Fake News}) and only real tweets (4th row, \textit{\#Only Real News}). For each of this four cases, the average and standard deviation have been calculated in both tables to show the fake/real tweet's rate and the uniformity of the users belonging to each of the aforementioned cases.

The results exposed in Table~\ref{tab:ActUsrsPol} are referred to the topic of political news. The amount of \textit{credulous} users that tweeted at least a fake news (1st row) is of 54,828, with publishing rate of 1.54 fake tweets on average and a standard deviation of 2.79. 138,113 \textit{credulous} users have been identified that published at least a real news (2nd row); and despite they are more numerous than previous case, their related tweeting rate (\textit{average}) is slightly lower (1.43) with an higher standard deviation (2.91). 

For what concern the amount of \textit{credulous} users publishing only fake/real news (3rd and 4th line), we can observe a small numerical decrease in quantity. There are 47,083 \textit{credulous} authors with, 1.40 tweets with fake news on average and the standard deviation is of 2.15. In the other case (4th row), 130,368 \textit{credulous} users have posted only tweets of real news, with almost the same average (1.37) as in its dual case but with an higher standard deviation (2.53).

As regards of \textit{not-credulous} users, we can notice that the authors posted at least a tweet containing a real news (2nd row) 
are 57,839, so almost 3 times than the users with at least a fake tweets (1st row), i.e.,19,525. The averages are similar in both cases, 1.98 for \textit{Real News} and 1.90 for \textit{Fake News}; the respective standard deviations are high too, 5.05 (1st row) and 10.22(2nd row). By observing those authors tweeted only fake/real news (3rd and 4th line), we can observe a similar trend to the same case of \textit{credulous}, but with a much lower level of participation. In fact, despite the \textit{not-credulous} authors posted only tweets of real news (51,480) are more than the ones published only fake tweets (13,526), 
these latter are only a third of the number of credulous users who only publish fake news. Furthermore, the \textit{not-credulous} users' tweeting rate of real news (\textit{average}) is also higher than the one referred to who is posting only fake news.

For sake of comparison to human accounts, in Table~\ref{tab:ActUsrsPol} are also reported the results related to \textit{bots}. We can see a certain disparity between the number of bots tweeting \textit{Fake News} (i.e, 9,622 as indicated in 1st row) and \textit{Real News} (i.e., 45,924 as indicated in 2nd row). On the other hand, by comparing the previous data with the ones related to bots authoring \textit{only} fake/real news (3rd and 4t line), we can see: (i) a reduction equal to more than 2 times about bots tweeting \textit{only} real news (17,515 in 4th line w.r.t the case in the 2nd line), and (ii) a little reduction for what concern the amount of bots sharing only fake news in their tweets (7,658 in 3rd line, w.r.t. 1st line). The tweeting averages per bots are upper than the human's cases; and the standard deviations have higher values when referred to real news (9.11) and \textit{only real news} (8.80). 

The outcomes concerning to gossip topic are presented in Table~\ref{tab:ActUsrsGoss}. Starting by the 1st macro-column (headed, \textit{credulous}), we can see a big amount of authors having at least one fake-news tweet (147,158), and this numerical superiority occurs even when we count the ones tweeting only fake news (139,187). Conversely, about the authors of real news' tweets, 39,528 credulous users have at least one and 31,557 published \textit{only} real tweets. 
By looking the details corresponding to the category of \textit{not-credulous} users, we can see a good downsizing of fake news' authors. Precisely, there are 56,451 users that tweetted at least one fake news, and 49,351 users that published only fake news in their tweets. Moreover, we observed the same decreasing trends also in both cases of real news published by \textit{not-credulous} authors. In particular, the authors published at least one real news are 19,047 (2nd row), whereas the ones published only real news are 11,947. 

Lastly, as in Table~\ref{tab:ActUsrsGoss}, we conclude by reporting the numerical details referred to the automated accounts (\textit{bot}, 3rd macro-column). We can summarize by saying that, the tweeting bots of at least a fake news are 25,818 and those ones published only fake are 19,540; more than the ones publishing real news, which are 14,620 (2nd row) and 8,342 (4th row). 
With regard to the average and standard deviation, the values are very high compared to the other lines, especially on standard deviation. In general, the averages of the bots is higher than both \textit{credulous} and \textit{not-credulous} users, regardless the news' veracity in their tweets.

From these results we can derive very interesting findings that will be exposed and discussed in the following Section~\ref{sec:5Disc}.

\section{Discussion}
\label{sec:5Disc}
The experimental results described in the previous section, 
shed light on 
the connections between fake news and \textit{credulous} users, and on the extent the latter are involved in spreading fake  news on Twitter. 
Here we provide some considerations about the experimental results along the three perspectives introduced in the Section~\ref{subsec:invTar}. Specifically, we consider 
\begin{itemize}
\item \textit{News}: the  differences  between  the  number  of  fake and  real  news  spread  by credulous, not-credulous and bots. 
\item \textit{Tweets}: the number of tweets containing either fake or real news posted by bots and humans and credulous and not-credulous. 
\item \textit{Activities}: the number of  users  tweet at  least  one fake  news,  at  least  one  real  news, and only fake or  real news.
\end{itemize}
 In addition, we provide some recommendation concerned with online monitoring of \textit{credulous} users.

\subsection{News}
The different behaviour of \textit{credulous} and \textit{not-credulous} users relatively to fake news, is made manifest in Table~\ref{tab:TopicXUsers}.
We can see that the tweets produced by \textit{credulous} users cover 95\% of the \textit{retrieved} fake news concerned with politics, 
while those produced by \textit{not-credulous} cover 92\%. Instead, for non fake news, \textit{credulous} users ``talk'' about the 91\% of the total \textit{retrieved}, which is almost the same of \textit{not-credulous} users. 

When considering gossiping news, the situation is a bit different. Specifically, on the one hand, the percentage of fake news covered by the \textit{credulous} users' tweets is 80\%, while \textit{not-credulous} users' tweets  93\% (
see Table~\ref{tab:Data}). 
On the other hand,
 \textit{not-credulous} users ``talk'' of 86\% of the \textit{retrieved} true news, while the coverage of \textit{credulous} users is of 92\%. This may seem counter-intuitive, but we have to stress the numerical imbalance between the amount of fake (more than 5k) and real gossip news (more than 15k). Moreover, in Table~\ref{tab:Data}, we consider as covered any news (re)twitted just once. In fact, considering the amount of tweets (Table~\ref{tab:TwGossip}) can be more informative.

%


From this first perspective, we can say that 
news do catch credulous users' interest, especially when fake content is concerned (80\% gossips vs. 95\% politics). 
In future work, we plan to investigate along new directions, by considering other topics such as technology or medicine.

\subsection{Tweets}
Regardless of news' veracity, from Table~\ref{tab:TwPolitic}, the low number of tweets made by bots compared to human-operated accounts attracts attention.
This is mainly due to the disproportion between the amount of bots and humans (see Table~\ref{tab:botDetector}), that confirms the statement in~\cite{Varol17} where the percentage of bots in Twitter is estimated to range from 9\% to 15\%; in our case is 10,52\%.
By looking to the human-operated accounts, we can see that the tweets authored by \textit{credulous} users are always more then those from the accounts classified as \textit{not-credulous}. 
This was 
expected about fake news; however, it was unexpected for the other news. Indeed, 
it would be a bit extreme to expect that \textit{credulous} users are active exclusively on fake news; however, it has also be taken into account that the amount of retrieved tweets of real news is more than three times the amount of fake ones. This can be observed by considering the 4th row in Table~\ref{tab:TwPolitic} (by comparing the columns headed \textit{FN} and \textit{RN}); we think that looking to the values in 3th (\textit{RN$_{rnd}$}) and 4th (\textit{RN$^*_{rnd}$}) columns would lead to a fairer comparison.
Concerning these two columns (for real tweets), 
it appears that \textit{credulous} users are authors of fewer real news tweets than fake ones, differently from \textit{not-credulous} users.

As already mentioned in Section~\ref{sec:4ExpRes}, also for gossips (Table~\ref{tab:TwGossip}), there is a downward trend between the amount of fake news tweets and real news ones for both categories of human-operated accounts.
In this case,
the situation is peculiar because the number of \textit{RN} is just less than \textit{FN} one; hence, considering the values related to the reduced set of real tweets (i.e., the column headed \textit{RN$_{rnd}$} and \textit{RN$^*_{rnd}$}) is somehow pointless.
%
The fact that the amount of \textit{not-credulous} users' real tweets does not overcome the number of the fake ones,
can be justified by a strong bot's authorship. In fact, the number of tweets done by bots on real news exceeds the number of its fake counterpart by more than 4 times (3 times if we consider the restricted sets \textit{RN$_{rnd}$} and \textit{RN$^*_{rnd}$}). 
A further motivation may be due to the fact that 
 gossip is a more attractive topic than politic 
and with a potential larger audience. In addition, traditional mass media (e.g., television, radio, newspapers, etc.) can be used to check ``veracity'' of political news, while it is more difficult to do the same for gossips.
However, we want just to emphasize the threat of the fake news, and even in this case the \textit{credulous} users ``win'' for number of fake tweets.

\subsection{Activities}
Further findings can be derived by observing the users' numerical participation, in each news' topic (Tables~\ref{tab:ActUsrsPol} and~\ref{tab:ActUsrsGoss}). It is worth to say that, in such tables, when we talk on the numbers related to real and fake news, we refer to the respective total set (\textit{FN} and \textit{RN}).  

Concerning political topics (Table~\ref{tab:ActUsrsPol}), we can claim that, despite there are more \textit{credulous} users publishing at least a real news, the related tweet's average is lower than the case when at least a fake news has been posted. It is also important to consider that the number of tweets of real political news is more than 410k, almost 3 times the amount of fake ones. Conversely, the (numerical) presence of \textit{not-credulous} users tweetting fake news is limited.

In the gossip's case (Table~\ref{tab:ActUsrsGoss}), we can see that, contrarily to what observed for the quantitative analysis of tweets (Table~\ref{tab:TwGossip}), the number of \textit{credulous} users who publish fake news is much higher than those who publish real ones.
Furthermore, the fact that the number of tweets with real news posted by \textit{credulous} users is relatively high suggests that they
tweet at greater intensity. And this even explains the unexpected high number (in Table~\ref{tab:TwGossip}) of real-tweets published by credulous users. An additional confirmation of this fact is given by observing just the authors of \textit{only} real or \textit{only} fake news. In this case, the amount of credulous users' that tweet real news is much lower than its fake equivalent. On the opposite, 
for \textit{not-credulous} users, we see a smaller presence in fake news cases; and, despite their number overcomes that of \textit{not-credulous} users that diffuse real news, the proportionality is significantly lower but with a higher average of real news's tweets (\textit{\#Real News $\geq$1}) authorship.

Although the 
bots have been included just for comparative purposes and the study of their involvement in fake news dissemination is not the main concern of our work, we can notice that 
they have the highest tweet's publishing rate compared to both categories of human-operated accounts. Although, from the study, 
their involvement in fake news spreading does not numerically emerge, we do think that they are among the main actors in this malicious activity and other literature provides some evidence. However, also in our study their malicious activity can this can be indirectly noticed by considering the average values of their fake-tweeting rate, which is higher than those of both categories of human-operated accounts.

To conclude this section, we can resume our findings in the following statements: 
\begin{enumerate}
\item Topics (politics vs gossips) play a key role in determining the amount of fake news published and circulating in social media, due to the size of potential audience and to the number of methods available for fact-checking; 
\item For both topics, \textit{credulous} users do spread a higher amount of fake news than \textit{not-credulous} users, but they do not publish only fake contents; 
\item A high number of \textit{credulous} users has a strong involvement in posting fake news; 
\item Automated accounts, which in general have a higher rate of tweets publication than humans, exhibit a good uniformity between fake and real content published.
\end{enumerate}

\subsection{Monitoring recommendations}
Despite  bots' tireless activity, the findings of this work shows active humans' participation in spreading fake news.
A possible way of taking advantage of our work is the design of a self-adaptive and evolving system that, by focusing on the data stream published by \textit{credulous} users' activity, carefully inspects what they publish in OSM, with targeted fact-checking.
The tool could be based on processes that intelligently exploit the data stream coming from credulous users as soon as these publish content on their dashboards. Then, by using text mining and/or NLP techniques, the tool could analyze such contents (tweets, retweets, replies, mentions, \ldots) in real-time, and, of course, the reliability of the source. This would considerably reduce the set of human users to scrutiny and, indirectly, the number of tweets, to perform targeted fact-checking. 
%
This tool could further evolve by targeting credulous more efficiently in order to further narrow the group of \textit{credulous} users to analyze. For instance, by considering content production's rate of \textit{credulous} users (to pay more attention to the more active ones), or the number of followers.

Such a tool, could be used by OSM administrators to hold up some of credulous users' activities (e.g., content re-posting) in order to slow down the propagation of malicious information.

\section{Conclusion}
\label{sec:6Conc}
Nowadays Online Social Media are very important and the  channel of information, preferred by people, especially by youngsters, to traditional media, like newspapers, radio and television. Their pervasiveness, favoured by the widespread use of mobile devices, and people's compulsive check of their social profiles has enabled faster news dissemination and a wider audience.
This has brought one of the bigger problem of our time, misinformation. The absence of any control of the news published on OSM, sort of common in newspapers, has stimulated production and diffusion of fake news. In many cases, this is done using automated accounts, called bots, which actively interact with people to induce bias in their opinion, generate misconception, incite hate-speech, etc. Current approaches, which effectively counter malicious bots activities, are based on their detection and removal from OSM. But whenever a bot is removed, it is easy to introduce new ones that are able to deceive detectors, giving rise to an arms race between botnet masters and OSM's administrators.

Inspired and stimulated by recent literature, that stresses the susceptibility of human-operated accounts to activities of malicious bots, we have studied the relationship between fake news and so called \textit{credulous} Twitter users, i.e., human-operated accounts following a high percentage of bots over their social contacts. 

Starting by a publicly available dataset of fake and real news (concerned with politics and gossips) and by using bots and credulous detectors, we provided evidence of the actual involvement of \textit{credulous} users in the diffusion/production of fake news. The experimental results showed that, regardless of the news topic, \textit{credulous} users tweet a larger number of fake news than not-credulous one. Although this superiority is also confirmed for the number of tweets containing real news, it is worth saying that:  (i) real news are harmless and (ii) the number of such tweets is still lower than those containing fake news. Furthermore, we observed the numerical participation of users, counting how many of each category have tweeted fake and/or real news; in this case we noticed a discordance depending on the news's topic. 
Credulous who have tweeted true political news, are in greater numbers than those who have published fake news, unlike the case of gossip news. But either way, the number of \textit{credulous} users tweeted fake news is always higher than the number of not-credulous users.

Because of this, we are pretty sure about the contribution of credulous detection techniques for improving fake news detectors; they would make it possible to focus on the content posted by credulous users using NLP and text mining techniques. We believe that the study of this category of users can help researchers to better understand misinformation and users' polarization phenomenons and can give an extra edge to fight the propagation of fake news.
It is worth to notice that the application of this kind of approach, that focuses on credulous users, is dependent from the classification performance of the adopted bot and credulous detectors.


Among the possible future work research directions, we plan to check if the findings of this work are confirmed by using other fake news datasets and apply fake news detection approaches (based on NLP and content inspection) to credulous users' tweets.


\section{Acknowledgment}
The authors acknowledge the support of the IT division at LNGS in providing the computing resources (ULITE) used to perform the reported experiments.

This work has been partially supported by the European Union’s Horizon 2020 program (grant agreement No. 830892, SPARTA) and by IMT School for Advanced Studies: Integrated Activity Project TOFFEe ‘TOols for Fighting FakEs’.
Finally, AB would like to thank Emilio Cruciani, Luca Di Stefano and Aline Uwimbabazi for discussions and suggestions.

\balance
\bibliography{EAIS20}
\bibliographystyle{IEEEtran}
\end{document}